\documentclass[usenatbib]{mn2e}
\usepackage{graphicx}
\usepackage{array}
\usepackage{multirow}
\usepackage[font=small,format=plain,labelfont=bf,up,textfont=normal,up]{caption}
\usepackage{float}
\usepackage{lscape}
\usepackage[normalem]{ulem}

\title[Polarization of Type 1 AGNs]{Spectropolarimetry of Seyfert 1 galaxies
with equatorial scattering: Black hole
masses and Broad Line Region characteristics}

\author[Afanasiev et al.]{V. L. Afanasiev$^{1}$\thanks{E-mail: vafan@sao.ru},
 L. \v C. Popovi\'c$^{2,3}$,  A.I. Shapovalova$^{1}$\\
$^1$ Special Astrophysical Observatory of the Russian
Nizhnij Arkhyz, Karachaevo-Cherkesia 369167, Russia \\
$^2$ Astronomical Observatory, Volgina 7, 11060 Belgrade 74, Serbia \\
$^3$ Department of Astronomy, Faculty of Mathematics, University
of Belgrade, Studentski trg 16, 11000 Belgrade, Serbia\\
 }

\begin{document}

\maketitle

\label{firstpage}

\begin{abstract}

Here we present the spectropolarimetric observations of a sample of  30 Type 1 AGNs and an analysis of the observed polarization in these AGNs. The observations have been performed with the 6-meter telescope of SAO RAS using  the modified SCORPIO-2 spectropolarimeter. We measured the Stokes parameters for the continuum and the broad H$\alpha$ line and obtained the values  of polarization degree and the angle of polarization.
We found that equatorial scattering is dominant polarization mechanism in the sample, that  allows us to use the observed polarization in the broad lines for determination of  the central black hole (BH) masses and characteristics (the inclination and emissivity) of the Broad Line Region (BLR). We demonstrated that the recently proposed method of \cite{ap15} for BH mass measurement gives accurate BH masses which are in a good correlation with the  stellar velocity dispersion, and consequently the masses determined by the polarization method can be used with calibration purposes. Additionally we found that the BLR in the sample of 30 AGN  has an averaged inclination of $35^\circ\pm9^\circ$ (mostly between 20 and 40 degrees)  and emissivity  $\alpha\sim -0.57$ that is more flat than one expected for the  classical accretion disc $\alpha\sim -0.75$.

\end{abstract}


\section{Introduction}

The unified model of active galactic nuclei (AGN)  established
after detection of a  polarized broad  H$\alpha$  line component in the spectrum of the   Seyfert 2 (Sy2) galaxy NGC 1068 \citep{am95}.
This indicates that the nature of  Seyfert 1 (Sy 1) and  Seyfert 2 galaxies is probably the same, but  the emission from  Sy 2  galaxies is blocked by the  torus  \citep{an93}. The polarization in the continuum of Sy 2 is due to polar scattering, but the broad emission component observed in some Sy 2 galaxies is coming from
the equatorial scattering, where emitted light from the Broad Line Region (BLR) is scattered on the inner part of the torus. Moreover, many of Sy 2 galaxies show the polarized broad lines in their spectra \citep[see e.g.][etc.]{mg90,tr92,yo96,he97,mo00,tr01,kis02,ki02,tr03,lu04,ra16}.

This discovery establishes the spectropolarimetry as a powerful tool for investigations of the AGN nature,
especially in the case of the Type 1 AGNs with prominent broad emission lines (BELs)
emitted by a BLR that  is supposed to be
very close to the central black hole (BH). The BLR has a geometry which
is probably driven by the central BH, and this reflects the
broad line shapes and parameters that can be used for estimation of the central BH mass \citep[see e.g.][for a review]{pet14}.
However, there are several problems in using the broad spectral lines for BH mass determination, one of them is that
the BLR is very small in size  (from several 10s  to several 100s light days), with an angular dimension $<10^{-5}$ arcsec for nearest AGNs \citep[][]{be15}
that cannot be resolved with the most powerful telescopes. Consequently, the BLR properties are far from being well characterized, and some of these properties have a strong impact on black hole mass determination. For example, the BLR inclination and geometry \citep[see][]{col06} may strongly affect the broad-line profiles, especially Full Width at Half Maximum (FWHM)  used in  the reverberation method. The reverberation method depends   on the FWHM, and virial factor
$f$ - that strongly depends on the BLR inclination \citep[see][]{pet14}. The uncertainties of   the $f$ and FWHM values can cause  large uncertainties in mass determination by the reverberation method. Therefore, it is very important to investigate the BLR  nature, and spectroscopy in combination with the measured polarization  the BELs that can give us some additional information about it \citep[][]{gg13}.

The BLR is supposed to have flattened shape, and the BLR emitting gas
is probably following a Keplerian-like motion.
It is assumed that the BLR has a clumpy
 structure, i.e. that it is composed from  a number of clouds
which radiate isotropically, consequently one
cannot expect  polarization due to the radiative transfer in the BLR. However,  the
polarization in the BELs can be due to the scattering of BEL light in the inner part
of the torus, i.e. the so-called equatorial scattering \citep[][]{sm04,sm05,af14,af15}.

This mechanism of polarization can provide more information about the BLR geometry, and, as it has been shown recently
by \cite{ap15} the polarization in BELs can be used for the mass measurement of the central BH in AGNs.
The mass determination by polarization method does not depend on the BLR inclination \citep[see][]{ap15}, and  small velocity
outflows/inflows do not affect
significantly the estimated BH masses  \citep[see][for more detailed discussion]{sav17}. Moreover,  opposite to with the reverberation method, where  the BLR virialization is assumed {\it a priori}, in the spectropolarimetric method the  Keplerian motion can be checked by using the relationship between the polarization angle and velocities across the broad line profile \citep[][]{ap15}. All these stated above is in a favor for using spectropolarimetric method to measure BH masses in the Type 1 AGNs.

Moreover, as we will demonstrate in the paper later,
the polarization in the broad emission lines can be used for the BLR inclination estimates.

The line profile reflects the characteristics of the velocity field in  BLR, and consequently polarization across the broad line profiles can be used for investigation of the  BLR structure \citep{ma98,sm02,sm04,sm05,gg07,af14,af15}. We observe a sample of Type 1
AGNs in order to investigate the BLR nature (inclination and emissivity) and measure the BH masses.

In this paper we present new observations of a sample of
30 Type 1 AGNs with 6-m BTA telescope, which some of them have been observed in
Spectropolarimetric monitoring campaign of AGN \citep{af11,af14,af15}. The aim of this paper is to discuss the
spectropolarimetric characteristics of Type 1 AGNs in the continuum and BELs. Special attention is paid to the new way of usingpolarization in the broad lines to find some characteristics of BLRs, as e.g. inclination and emissivity.
  The paper is organized as following: in \S 2 we describe observations and
data reduction, in \S 3 the results  are given and analyzed  in \S 4.
Finally,  in
\S 5 we present a short discussion and outline our conclusions.

\section{Observations and data reduction}

The observations have been performed with 6-m telescope of SAO RAS.
We used the modified SCORPIO-2 spectrograph \citep{am11,ar14} in the spectropolarimetry mode . The log of
observations is presented in Table \ref{obs-log},
where the name of objects,  the redshift, the date of observations, the exposure and  the seeing are give.
 A double Wollaston prism - WOLL2 \citep{oli97} is used as a polarization analyzer.
In this analyzer the rays that enter into two Wollastons which have crystal
axis at $45^\circ$, and emerge at 4 different angles with relative
intensities depending on the polarization angle and degree
of the input light. The four images  correspond therefore to measurements performed with polarizers at 0, 90, 45 and 135 degrees. The obtained couple images of an object are distributed along the slit using  achromatic wedges. In each exposition the four spectra have been recorded simultaneously  in different polarization planes. Thus, we can measure three Stokes parameters I, Q and U, from only one observation, unlike  a single Wollaston analyzer usage, demanding to obtain four exposures in different  angles of polarization plane.

\begin{table}
\caption[]{Summary log of observations. From the left to the right are given: name of object,
redshift,  date of observations, exposure in sec and seeing in arcsec.}
\label{obs-log}
\begin{tabular}{lcccc}
\hline

Object  &z&Date&Exposure&Seeing\\
        & & dd/mm/yy    & (sec)& (arcsec)\\
\hline
Mkn335&0.026&09/11/13 &2400&1.1\\
Mkn1501&0.098&20/11/14 &3600&1.2\\
Mkn1148&0.064&20/11/14 &2880&1.2\\
1Zw1&0.059&22/11/14 &2400&2.5\\
IRAS03450+0055&0.031&20/10/14 &3600&2.0\\
3C120&0.033&06/11/13 &3600&2.9\\
Akn120&0.032&24/03/14 &3600&1.5\\
MCG+08-11-011&0.020&06/11/15 &3600&1.0\\
Mkn6&0.019&25/02/13 &5400&2.5\\
Mkn79&0.022&07/12/15 &3600&1.1\\
PG0844+349&0.064&07/04/16 &3600&1.3\\
Mkn704&0.029&23/11/14 &3420&2.5\\
Mkn110&0.035&12/11/14 &2400&1.1\\
NGC3227&0.004&10/12/15 &1800&1.0\\
NGC4051&0.002&24/03/14 &1800&1.5\\
NGC4151&0.003&23/05/14 &1200&1.5\\
3C273&0.158&25/03/14 &1440&1.5\\
NGC4593&0.009&18/03/15 &1800&2.0\\
Mkn231&0.042&18/03/15 &1800&1.8\\
IRAS13349+2438&0.108&23/05/14 &2400&1.3\\
Mkn668&0.077&24/03/15 &5400&2.5\\
NGC5548&0.017&25/03/14 &2400&1.5\\
Mkn817&0.031&29/05/14 &3600&1.2\\
Mkn841&0.036&29/05/14 &4200&1.1\\
Mkn876&0.129&26/03/15 &3900&2.0\\
PG1700+518&0.292&05/04/16 &3600&1.7\\
3C390.3&0.056&09/05/14 &3600&1.0\\
Mkn509&0.034&21/10/14 &3360&1.9\\
Mkn304&0.066&09/11/13 &3600&1.1\\
3C445&0.056&05/11/13 &3600&1.5\\
\hline
\end{tabular}

\end{table}

\begin{figure*}
\centering
\includegraphics[width=17.5cm]{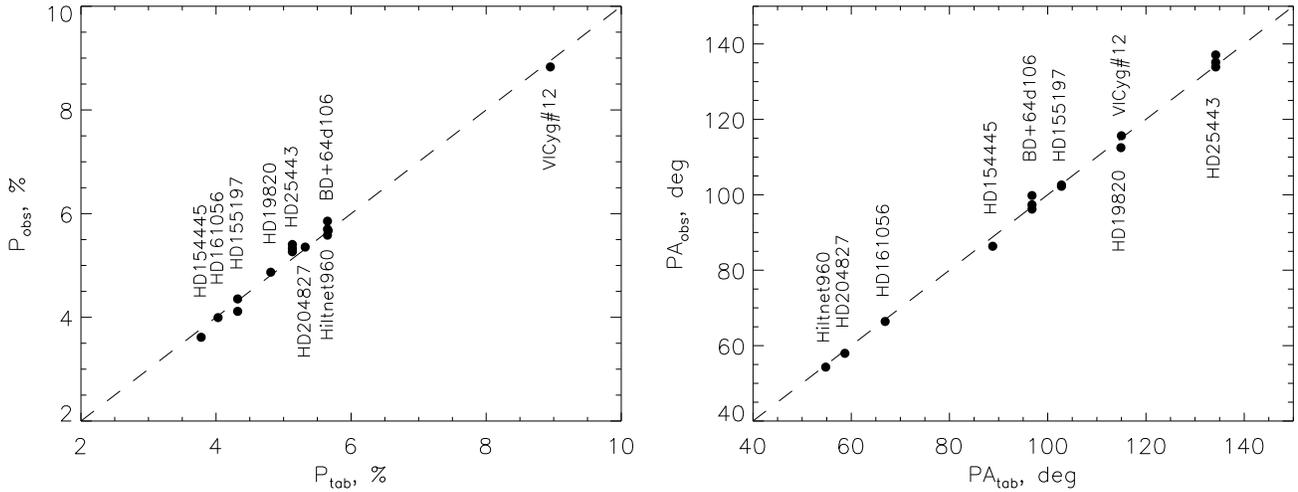}
\caption{Comparison of  observed  values { in V(5500\AA) waveband} and  ones given in literature of the polarization degree (left) and polarization angle
(right) for polarization standard stars.}
\label{compare-stars}
\end{figure*}

 In this mode, the slit length is 1 arcmin. The width of the slit,  depending  the  seeing, was  between 1.5 and 2 arcsec. The spectral resolution is caused by the
width of the slit and used spectral grating. Typical spectral resolution was 8-11 \AA. We used two gratings -  VPHG940 (spectral coverage 4000-7500 \AA\AA) and  VPHG1026 (spectral coverage 5800-9500 \AA\AA)\footnote{The parameters of these spectral gratings can be found at https://www.sao.ru/hq/lsfvo/devices/scorpio-2/grisms\_eng.html}.
Unlike conventional diffraction gratings in the case of VPHG, the difference between polarization in directions parallel and perpendicular to the direction of dispersion is small (about 5\%).
This makes VPHG very suitable for spectrophotometric observations.

For the WOLL2 analyzer three Stokes parameters have been found using following relations:

$$I(\lambda)=I_0(\lambda)+I_{90}(\lambda)K_Q(\lambda)+I_{45}(\lambda)+I_{135}(\lambda)K_U(\lambda)\eqno(1)$$

$$Q(\lambda)={I_{0}(\lambda)-I_{90}(\lambda)K_Q(\lambda)\over{I_{0}(\lambda)+I_{90}(\lambda)K_Q(\lambda)}},\eqno(2)$$

$$U(\lambda)={I_{45}(\lambda)-I_{135}(\lambda)K_U(\lambda)\over{I_{45}(\lambda)+I_{135}(\lambda)K_U(\lambda)}},\eqno(3)$$
where $K_Q$ and $K_U$ are the instrumental parameters related to  the transmission of the polarization channels. Each channel has been corrected  for  spectral sensitivity of the device, which is  determined from observations of the polarization standards. The $I_0(\lambda),\ I_{90}(\lambda)),\ I_{45}(\lambda)$  and $I_{135}(\lambda)$ are the intensity in the four polarization angles.

The  degree of linear polarization $P(\lambda)$ and  polarization angle $\varphi(\lambda)$ have been obtained using following
relations:
$$P(\lambda)=\sqrt{Q(\lambda)^2+U(\lambda)^2}, \ \ \ \ \varphi(\lambda)={1\over 2}{\rm arctg}[U(\lambda)/Q(\lambda)]+\varphi_0,
\eqno(4)$$
where $\varphi_0$ is the zero point of polarization angle, which also has been determined by the observations  the polarization standards.

Observations for each object have been performed in a series of exposures (more than 5), with the integration aperture of 2$\times$10 arcsec. Before or after the object exposure, polarization standards and zero-polarization stars have been observed \citep[the standards are taken from][]{tu90,hs82}. The typical object exposures are  180-240 seconds that depends on the object brightness. This type of observations allowed us to have a  robust
statistical estimation of the observed values \citep[see][ for a detailed discussion]{af16}. The method of observations and data reduction  was described in more details by \cite{aa12}.
In the case when the galactic latitude was smaller than 25 degrees, we took into account the
interstellar polarization as an average polarization of several stars around the observed object, as it was described by
\cite{af14}. The accuracy of our observation is limited by the variation of the depolarization in the Earth atmosphere.

In Fig. \ref{compare-stars} we give our measurements of the polarization degree  and polarization
angle ($P_{obs}$ and $PA_{obs}$) for the standard polarization of
stars, and compare our measurements with those given in the literature \citep[$P_{tab}$ and $PA_{tab}$ taken from][]{sch92}.
The error of the measurements of polarization is  ($P_{tab}-P_{obs}$)~$\sim$ 0.2\%, and the
polarization angle around  ($PA_{tab}-PA_{obs}$)~$\sim$ 3 degrees.
\begin{figure}
\centering
\includegraphics[width=8.5cm]{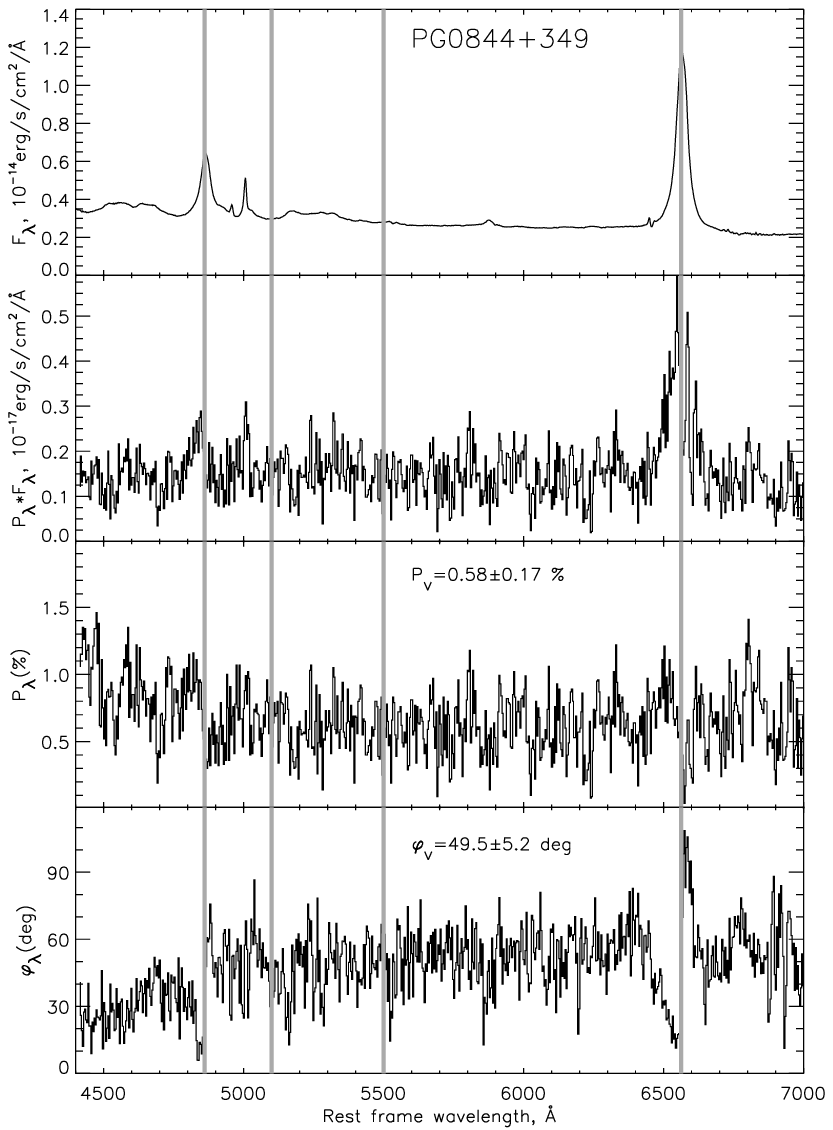}
\caption{Spectropolarimetric observations of quasar PG0844+349, from the top to the bottom: integral spectrum, polarized spectrum, degree of polarization and polarization angle. The vertical lines denote the position (from left to the right) of H$\beta$, continuum at $\lambda$5100 \AA\,  continuum at $\lambda$5500 \AA\  and H$\alpha$ line}.
\label{example}
\end{figure}

\begin{table*}
\begin{center}
\caption[]{The data for the continuum polarization. From the left to the right are given: object name, flux at 5100 \AA\  measured from our observations, unbiased degree of polarization in two wavebands $V(5500\AA)$ and $R(6500\AA)$, index $n$ of the wavelength dependent slope,
polarization angles for $V(5500\AA)$ and $R(6500\AA)$ wavebands, $P(ref)$ and $PA(ref)$  are the polaization degree and angle taken from reference given in column
'Ref' in  the V-band.}
 \label{cont-data}

\begin{tabular}{lccccccccccc}

\hline \hline
\\
Object&Flux(5100)&$P(V)$&$P(R)$& $n$&$\varphi(V)$&$\varphi(R)$&$P(ref)$& $PA(ref)$&Ref\\

\hline
Mkn335          & 7.53$\pm$0.54&0.41$\pm$0.14&0.36$\pm$0.06&-0.78$\pm$0.49&110.9$\pm$14.9& 98.9$\pm$5.9 &0.48$\pm$0.11&107.6$\pm$6.9& 1\\
Mkn1501         & 1.59$\pm$0.46&0.94$\pm$0.34&1.25$\pm$0.60& 1.71$\pm$0.90&140.9$\pm$16.2&153.4$\pm$12.6&                   &                   &       \\
Mkn1148         & 1.38$\pm$0.15&0.67$\pm$0.22&0.98$\pm$0.45& 2.28$\pm$0.85& 67.1$\pm$11.1& 81.0$\pm$16.2&0.81$\pm$0.28&  77$\pm$10&2\\
1Zw1            & 8.21$\pm$1.02&0.91$\pm$0.21&0.77$\pm$0.16&-1.00$\pm$0.44&146.7$\pm$ 8.4&149.6$\pm$ 6.5&0.70$\pm$0.05&  148$\pm$2&1\\
IRAS03450+0055  & 3.15$\pm$0.26&0.93$\pm$0.24&0.86$\pm$0.34&-0.47$\pm$0.72&115.9$\pm$ 7.8&119.5$\pm$11.0&&&\\
3C120           & 6.38$\pm$0.38&1.17$\pm$0.25&1.10$\pm$0.24&-0.37$\pm$0.44&111.4$\pm$ 6.6&103.5$\pm$7.9 &0.92$\pm$0.25&103.5$\pm$7.9&4\\
Akn120          & 7.91$\pm$0.44&0.59$\pm$0.17&0.70$\pm$0.29& 1.02$\pm$0.76& 78.7$\pm$ 6.9& 70.9$\pm$ 8.9&0.65$\pm$0.13&78.6$\pm$5.7&4\\
MCG+08-11+011   & 5.73$\pm$0.36&1.38$\pm$0.15&1.36$\pm$0.25&-0.09$\pm$0.33& 71.9$\pm$ 3.6& 76.6$\pm$ 5.9&1.69$\pm$0.46&76.4$\pm$19&4\\
Mkn6            & 8.96$\pm$0.63&0.71$\pm$0.20&0.78$\pm$0.16& 0.56$\pm$0.47&155.2$\pm$ 9.3&148.2$\pm$ 8.0&0.90$\pm$0.02&156.5$\pm$0.8&1\\
Mkn79           & 3.64$\pm$0.22&1.43$\pm$0.24&1.20$\pm$0.22&-1.05$\pm$0.36&  4.2$\pm$ 4.6& -0.2$\pm$11.5&1.34$\pm$0.19&0.4$\pm$16.2&4\\
PG0844+349      & 2.97$\pm$0.15&0.58$\pm$0.17&0.58$\pm$0.20& 0.00$\pm$0.66& 49.5$\pm$ 5.2& 49.1$\pm$ 9.8&&&\\
Mkn704          & 3.89$\pm$0.24&2.32$\pm$0.36&2.21$\pm$0.41&-0.29$\pm$0.36& 58.3$\pm$ 2.6& 56.8$\pm$ 5.3&2.01$\pm$0.21&68.1$\pm$5.1&3\\
Mkn110          & 3.69$\pm$0.28&0.55$\pm$0.16&0.38$\pm$0.23&-2.21$\pm$1.05& 19.2$\pm$ 0.5& 21.8$\pm$ 7.3&0.21$\pm$0.13&18.$\pm$15.&2\\
NGC3227         &10.32$\pm$1.97&1.25$\pm$0.42&0.96$\pm$0.29&-1.58$\pm$0.64&137.8$\pm$11.2&142.2$\pm$13.2&1.3$\pm$0.1&133$\pm$3.&5\\
NGC4151         &40.52$\pm$7.97&0.32$\pm$0.30&0.20$\pm$0.11&-2.81$\pm$2.78& 69.7$\pm$ 6.2& 68.2$\pm$ 6.3&0.26$\pm$0.08&62.8$\pm$8.4&4\\
NGC4051         &17.78$\pm$1.24&0.39$\pm$0.19&0.22$\pm$0.21&-3.43$\pm$1.67& 92.6$\pm$10.2& 98.2$\pm$ 3.4&0.87$\pm$0.04&78.0$\pm$0.8&4\\
3C273           &26.61$\pm$1.61&1.11$\pm$0.17&1.34$\pm$0.34& 1.13$\pm$0.45& 57.8$\pm$ 4.2& 61.3$\pm$ 3.9&0.87$\pm$0.11&65$\pm$3&8\\
NGC4593         &10.87$\pm$0.56&0.77$\pm$0.20&0.62$\pm$0.08&-1.30$\pm$0.37&106.9$\pm$ 9.6&101.6$\pm$10.8&0.57$\pm$0.16&109.5$\pm$11.&2\\
Mkn231          & 8.88$\pm$0.40&3.29$\pm$0.29&2.98$\pm$0.63&-0.59$\pm$0.36&118.4$\pm$ 3.3&118.8$\pm$ 6.8&2.87$\pm$0.08&95.1$\pm$1.&4\\
IRAS13349+2438  & 5.23$\pm$0.20&5.59$\pm$0.27&5.42$\pm$0.39&-0.18$\pm$0.13&125.5$\pm$ 0.9&126.2$\pm$ 3.7&5.70$\pm$0.12&124.4$\pm$0.7&9\\
Mkn668          & 1.87$\pm$0.18&1.06$\pm$0.22&0.91$\pm$0.32&-0.91$\pm$0.63&127.7$\pm$ 5.6&136.6$\pm$13.2&0.91$\pm$0.55&145$\pm$3&7\\
NGC5548         & 9.49$\pm$0.39&0.50$\pm$0.12&0.40$\pm$0.15&-1.34$\pm$0.68& 34.2$\pm$ 5.1& 34.7$\pm$ 5.7&0.72$\pm$0.10&33.5$\pm$3.9&2\\
Mkn817          & 3.86$\pm$0.15&0.71$\pm$0.15&0.57$\pm$0.21&-1.31$\pm$0.65& 76.4$\pm$ 7.7& 77.9$\pm$18.0&&&\\
Mkn841          & 5.30$\pm$0.08&0.78$\pm$0.33&1.13$\pm$0.38& 2.22$\pm$0.75&100.7$\pm$14.8&102.3$\pm$12.8&1.00$\pm$0.03&92.9$\pm$4.3&4\\
Mkn876          & 3.37$\pm$0.07&0.94$\pm$0.23&0.78$\pm$0.19&-1.12$\pm$0.49&113.8$\pm$ 9.4&111.5$\pm$13.6&0.81$\pm$0.04&110.5$\pm$3.7&1\\
PG1700+518      & 3.10$\pm$0.10&0.77$\pm$0.22&0.89$\pm$0.35& 0.87$\pm$0.73& 71.7$\pm$12.7& 53.0$\pm$15.5&0.54$\pm$0.10&56$\pm$5&10\\
3C390.3         & 1.69$\pm$0.08&0.77$\pm$0.08&0.91$\pm$0.37& 1.00$\pm$0.68&139.5$\pm$ 0.8&141.6$\pm$ 7.7&0.74$\pm$0.3 &146$\pm$5&6\\
Mkn509          & 7.15$\pm$0.28&0.68$\pm$0.15&0.52$\pm$0.34&-1.61$\pm$1.10&149.1$\pm$ 7.2&150.7$\pm$ 7.5&1.09$\pm$0.15&146.5$\pm$4.0&4\\
Mkn304          & 3.62$\pm$0.57&0.58$\pm$0.20&0.61$\pm$0.11& 0.30$\pm$0.50&131.1$\pm$11.3&130.2$\pm$ 6.0&0.98$\pm$0.14&136.6$\pm$4.2&4\\
3C445           & 1.18$\pm$0.06&2.86$\pm$0.55&1.98$\pm$0.58&-2.20$\pm$0.53&152.2$\pm$ 4.2&147.4$\pm$10.5&3.01$\pm$0.30&150$\pm$5&6\\

\hline
\end{tabular}
\\
{UNITS: Flux(5100) is given  in	$10^{-15}\rm erg\ cm^{-2}s^{-1}\AA^{-1}$, $P(V)$ and $P(R)$ in percents, $\varphi(V)$ and $\varphi(R)$ in degrees\\}
{ References :  1 - Smith et al. (2002),   2 - Berriman et al. (1990),  3 - Goodrich and Miller (1994),
4 - Martin et al. (1983),  5 - Schmidt and Miller (1985), 6 - Kay et al. (1999), 7 - Corbett  et al. (1998) ,
8 - Wills at al. (1992a), 9 - Wills  at al. (1992b) \\}

\end{center}
\end{table*}

 The observed spectra have been corrected to the system sensitivity function and extinction.
 The flux calibration (given in erg s$^{-1}$cm$^{-2}$A$^{-1}$) is performed by using
 the spectrophotometic standards (in this case we used the spectra of  G191b2b, BD+28d4655 and BD+33d2642 stars). The continuum at 5100 \AA\ and measurements of the polarization parameters from our spectra  are given in Table \ref{cont-data}.

The typical spectra are shown in Fig. \ref{example}, where
we present (from top to bottom) the unpolarized and polarized spectra, the degree of
polarization and the polarization angle
for quasar PG0844+349. The wavelengths are rescaled to the rest-frame. The vertical lines denote the position of the H$\beta$ and H$\alpha$ lines and the spectral region of
the unpolarized continuum measurement at 5100 \AA\ and polarized one at  5500 \AA.

\section{Results}

We analyzed our observations in order to give the polarization parameters in the continuum and broad lines.
As it can be seen in Fig. \ref{example} we can detect the H$\beta$ and H$\alpha$
broad lines in polarized light, however H$\beta$ is generally weaker and S/N ratio in
the polarized light is smaller than in H$\alpha$. Therefore, we will use the
polarized H$\alpha$ line to measure parameters in broad lines. Our measurements are
given in Tables \ref{cont-data} and \ref{tab3} and Figs. \ref{res1} -- \ref{res6}.
We separately discuss
the obtained polarization parameters in the continuum and lines.

\subsection{Polarization in the continuum}

 The estimation of the polarization in the continuum is presented  in Table \ref{cont-data},
where we give the object name (1st column), the observed fluxes   at  5100\AA ~in the aperture 2$\times$10 arcsec in units of $10^{-15}\rm erg\ cm^{-2}s^{-1}\AA^{-1}$ (2nd column). The   spectra of nucleus were obtained by subtracting the host galaxy spectrum from the observed one using the technique described by \citet{ar14}.
 Properties of the continuum polarization for each object  were defined as follows.
Assuming that $V(5500\AA)$ waveband covers the wavelength interval of $\lambda\lambda=5300-5700\AA$,
and $R(6500\AA)$  interval $\lambda\lambda=6300-6700\AA$ (in the rest frame). In Table \ref{cont-data} we offer the averaged values of polarization degree (column 3rd and 4 th) in these intervals (P(V) and P(R)) in percents.
In cases where the measured  polarization degree was
on the level of its error, the P(V) and P(R) values are biased. In Table \ref{cont-data}
we give unbiased values of the degree of polarization, which were calculated as \citep[see][]{sim85}:
$$P_{unbiased}=\sqrt{P^2-1.41\sigma_P^2}, $$ where $P$ is the measured polarization and
$\sigma_P$ is the corresponding error.
We also find the index $n$ \citep[that has been defined by][]{af11}
which indicates the dependence of polarization as $P\sim \lambda^n$. Index $n$, which represents the inclination of polarized continuum, was calculated  using the measured polarization in $V(5500\AA)$ and $R(6500\AA)$ wavebands as:
$$n={{\log P(V)-\log P(R)}\over \log(5500/6500)}.$$

\begin{figure*}
\centering
\includegraphics[width=16cm]{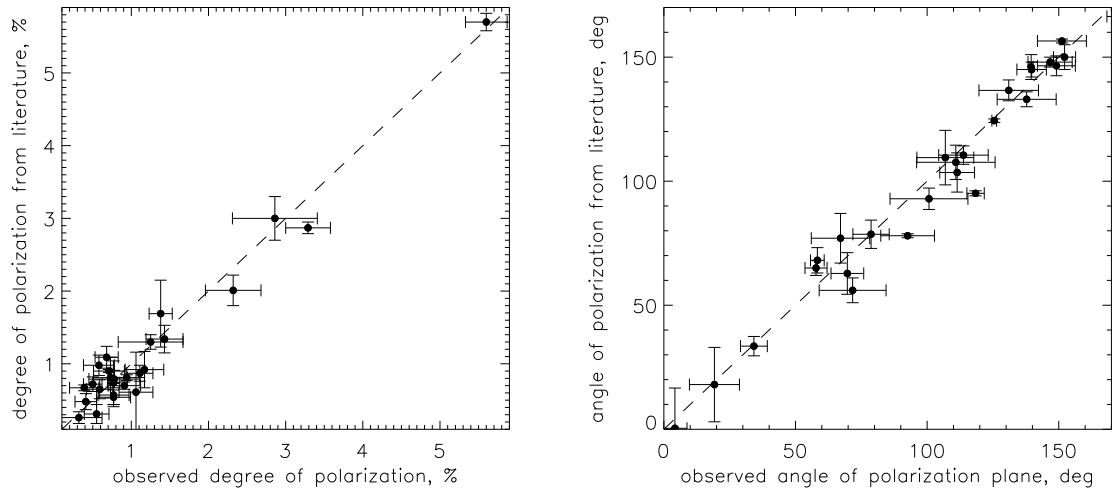}
\caption{Comparison of our measurements in V(5500\AA) waveband with ones given in literature:
 the degree of polarization (left) the angle (right)}
  \label{pol-AGN}
\end{figure*}

The index $n$
in the first approximation describes the depolarization of the radiation as a function of
wavelengths and can be used for the exploring of the polarization mechanisms. For example, if the observed polarization is caused by Thompson scattering due to radiation transfer then it does not depend from wavelength,while  if there is a wavelength dependence of polarization in the continuum, one has to suppose other mechanisms to explain it.
The index $n$ and polarization  angles for V- and R-bands are given in Table \ref{cont-data} --
columns 5th, 6th and 7th respectively. Several of AGNs from our sample,  polarization degrees and angles have been observed earlier in the optical V-band. We show  these data in Table \ref{cont-data}
(columns 8th and 9th, for polarization degree and angle, respectively) and in the
10th column we quote the corresponding reference.
In Fig. \ref{pol-AGN} we compare  our measurements and those
given in literature. As one can see in Fig. \ref{pol-AGN}, there are no systematic differences between our and data  taken from literature. The scatter between our data and data from literature  is probably due to measurement errors and possible variability of objects.

\subsection{Polarization in the broad lines}

As we mentioned above, we expect that a dominant polarization mechanism in the BLR light is the equatorial
scattering \citep[see][]{sm05,ap15}. In the case of the equatorial scattering, and
Keplerian-like motion of broad line emitting gas, the
polarization angle has specific shape \citep[see][]{af14,ap15}. The horizontal S-like shape of polarization
angle in H$\beta$ and H$\alpha$ can be detected  in all observed objects.
The spectral and polarization parameters around H$\alpha$ for each object were measured (see Figs.  \ref{res1}-\ref{res6}).

 In Figs.\ref{res1}-\ref{res6} we show
the  top to bottom: the H$\alpha$ line profile, the parameters $Q$ and
$U$ across the line profile, the degree of polarization ($P$) and
polarization angle ($\Delta\varphi$)  across the line profile

 The averaged parameters of $<Q>$,  $<U>$, $<P>$ and $<\varphi>$
measured in the continuum around  the H$\alpha$ line (R-band)
are given on the corresponding plot.

\subsubsection{Estimates of BH masses}

It was shown in \cite{af14} and
\cite{ap15} that in the case of the Keplerian-like motion (and equatorial scattering of the BLR
light) the velocities across the broad line and $\tan\varphi$ are connected as
(see Fig. \ref{angle-V}):

$$\log({V_i\over c})=a-b\cdot \log(\tan(\Delta\varphi_i)), \eqno(5)$$
where $c$ is the speed of light, and constant $a$   depends on the BH mass ($M_{BH}$) as

$$a=0.5\log\bigl({{GM_{BH} \cos^2(\theta)}\over{c^2R_{sc}}}\bigr). \eqno(6)$$
where $G$ is the gravitational constant, $R_{sc}$ is the distance of the
scattering region { from the central black hole and $\theta$ is the angle between the BLR disk and the scattering region. In the case of equatorial scattering, we can assume the $\theta\sim0$ \citep[i.e. $\cos^2\theta\sim 1$, see][]{ap15,sav17} and therefore, the BH  mass is independent from the inclination. $b$ is 0.5 for the case of the Keplerian motion.}

The equations above are  obtained  using the approach of  a simple
geometric model \citep[see][]{ap15}, but it is shown in \cite{sav17}
that this approximation gives good results.
 In the  paper  \cite{sav17}   validity of the method was explored using  numerical
simulations by applying the  modified STOKES code \citep{mar12}  showing that the observed rotation of the polarization plane depending on the velocity in the broad  emission lines
for the Keplerian motion in the BLR. The motion is well described by Eq. (5)
and obtained masses are weakly dependent on the geometry of the torus and the inclination
of the BLR.

\begin{landscape}
\begin{figure}
\centering
\includegraphics[width=24cm]{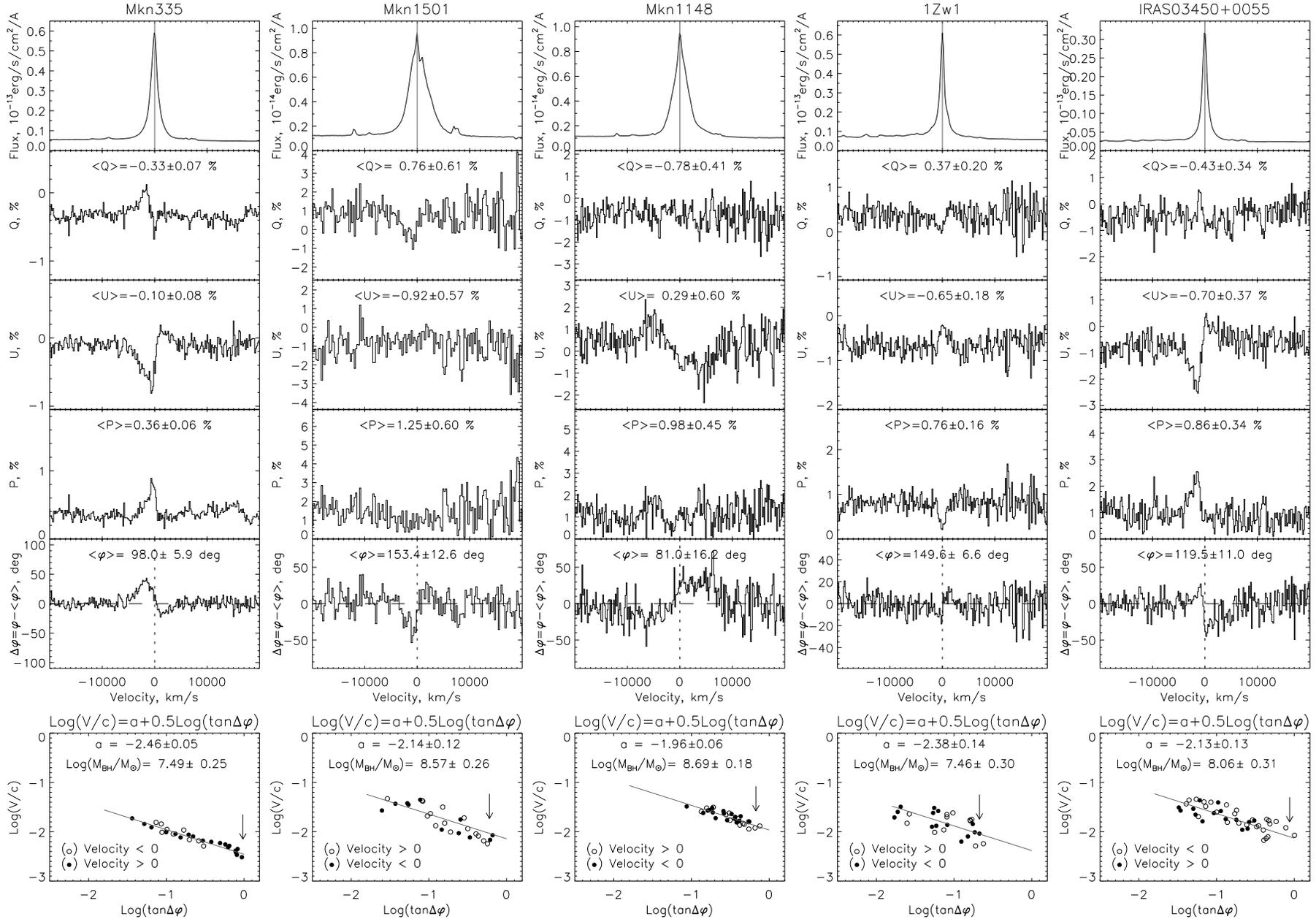}
\caption{The H$\alpha$ spectral region (1st panel), the shape of Stokes $Q$ and $U$ parameters
(2nd and 3rd panels, respectively) across the H$\alpha$ line profile, and corresponding
percent of polarization (P\%) and polarization angle - $\Delta\varphi$ (4th and 5th panels,
respectively). On bottom panels the relationship of
$\log(v/c)$ {\it v.s.} $\log\tan(\Delta\varphi)$ is shown.
From left to the right the (column) panels are for  Mkn 335, Mkn 1501, Mkn 1148, 1Zw1 and IRAS03450+0055.}
\label{res1}
\end{figure}
\end{landscape}

\begin{landscape}
\begin{figure}
\centering
\includegraphics[width=24cm]{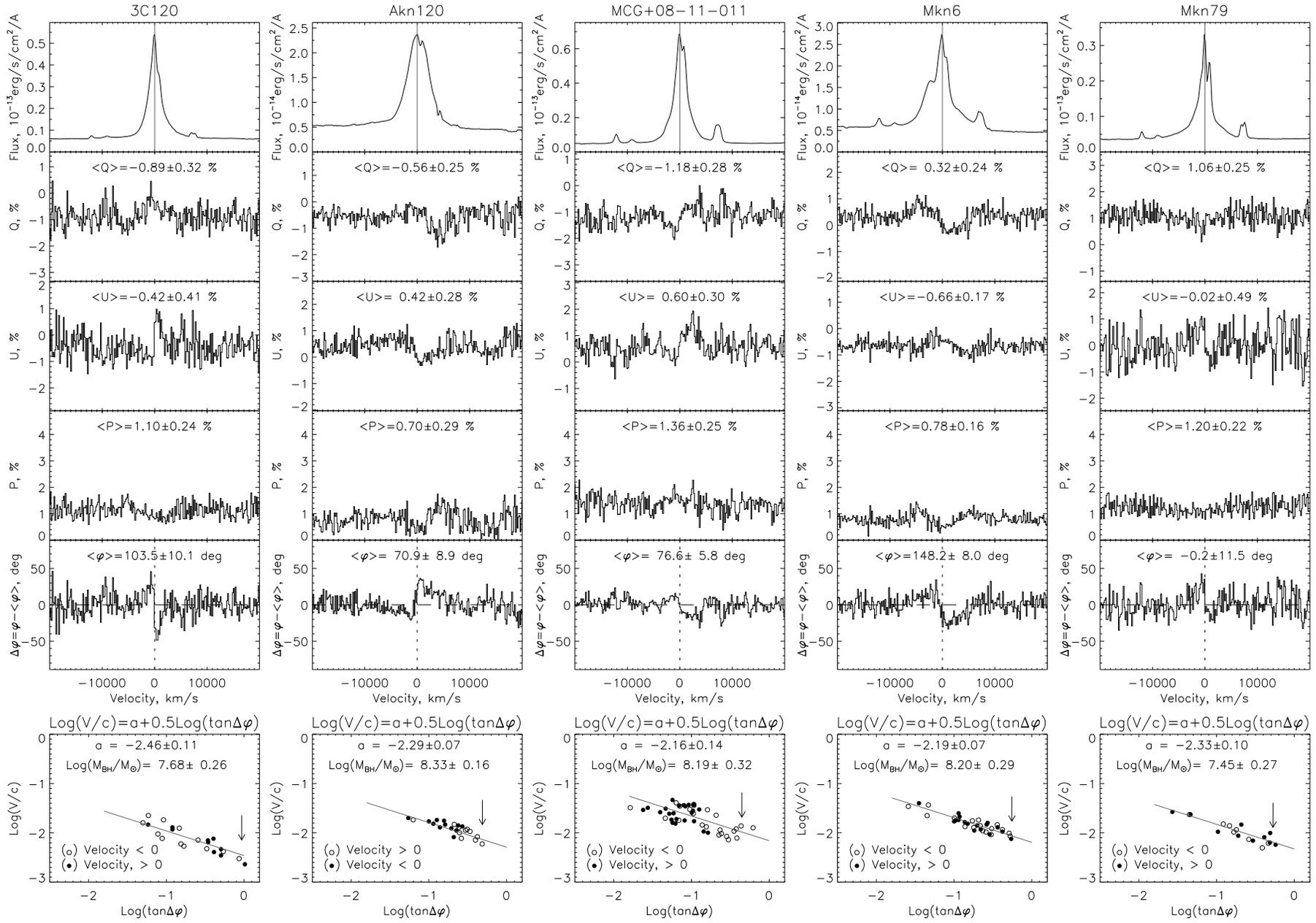}
\caption{The same as in Fig. \ref{res1}, but for 3C 120, Akn 120, MCG+08-11-011, Mkn 6 and Mkn 79.}
\label{res2}
\end{figure}
\end{landscape}

\begin{landscape}
\begin{figure}
\centering
\includegraphics[width=24cm]{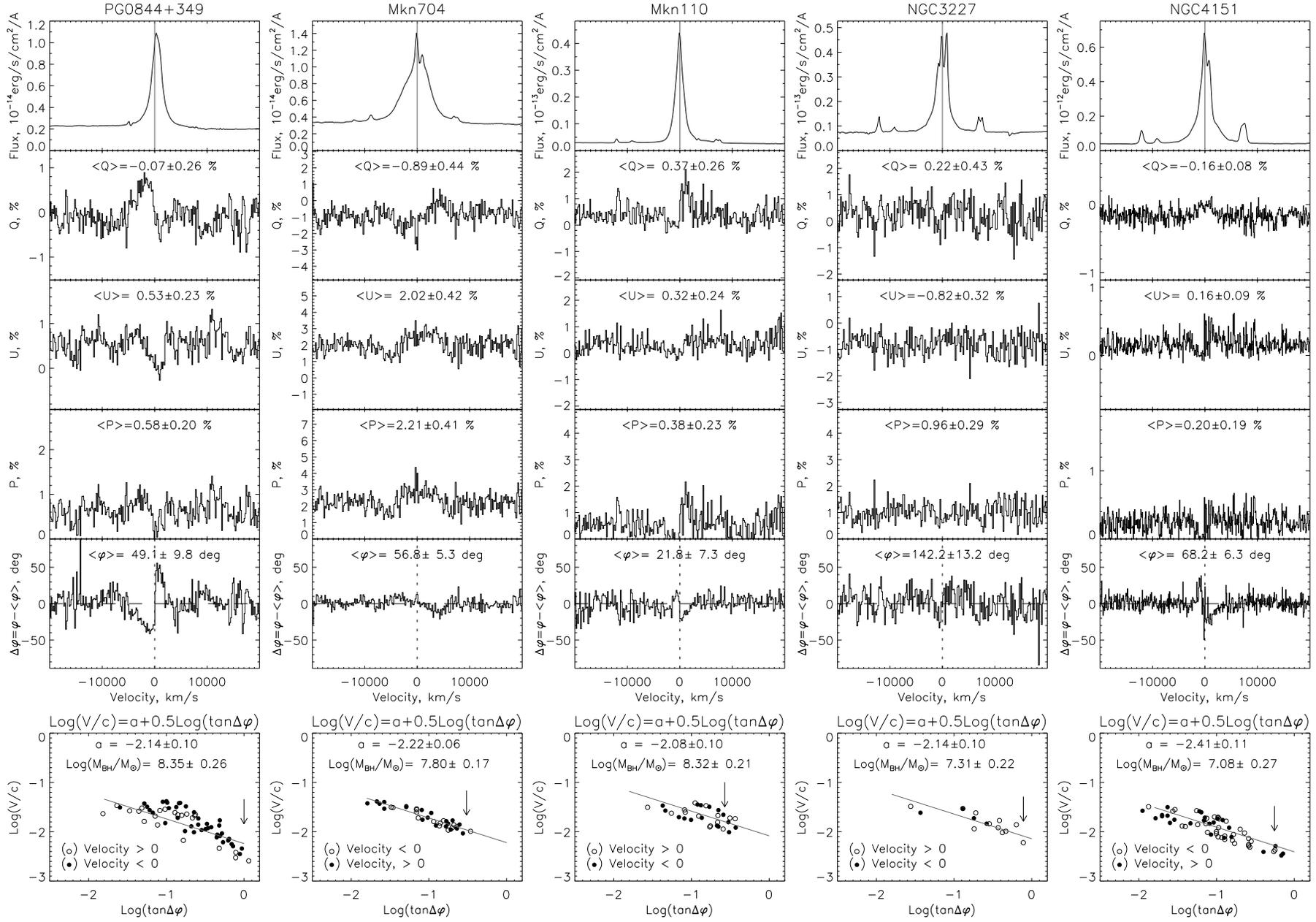}
\caption{The same as in Fig. \ref{res1}, but for PG 0844+349, Mkn 704, Mkn 110, NGC 3227 and
NGC 4151.}
 \label{res3}
\end{figure}
\end{landscape}

\begin{landscape}
\begin{figure}
\centering
\includegraphics[width=24cm]{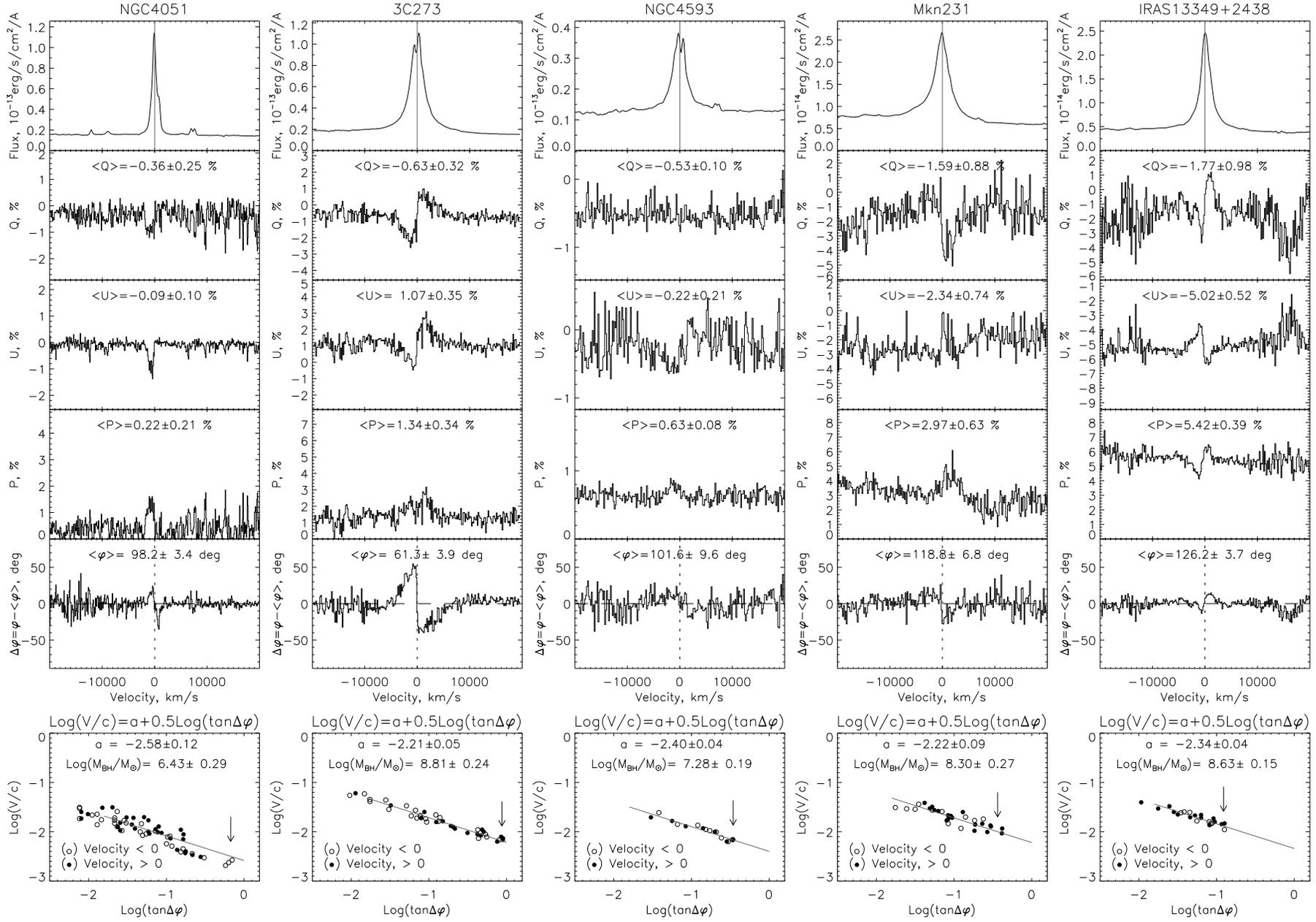}
\caption{The same as in Fig. \ref{res1}, but for NGC 4051, 3C 273, NGC 4593, Mkn 231 and
IRAS13349+2438.}
 \label{res4}
\end{figure}
\end{landscape}

\begin{landscape}
\begin{figure}
\centering
\includegraphics[width=24cm]{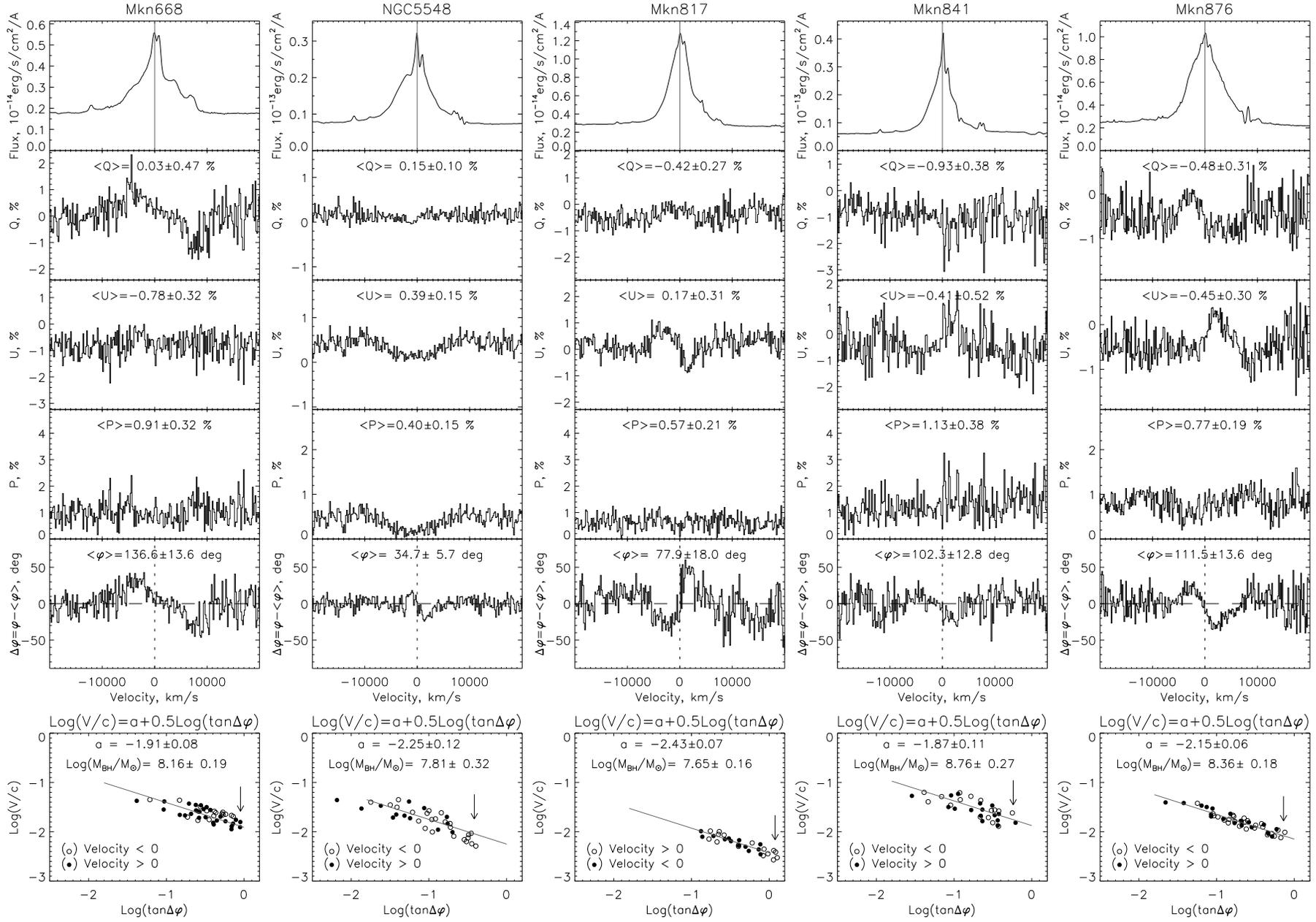}
\caption{The same as in Fig. \ref{res1}, but for Mkn 668, NGC 5548, Mkn 817, Mkn 841 and Mkn 876.}
 \label{res5}
\end{figure}
\end{landscape}

\begin{landscape}
\begin{figure}
\centering
\includegraphics[width=24cm]{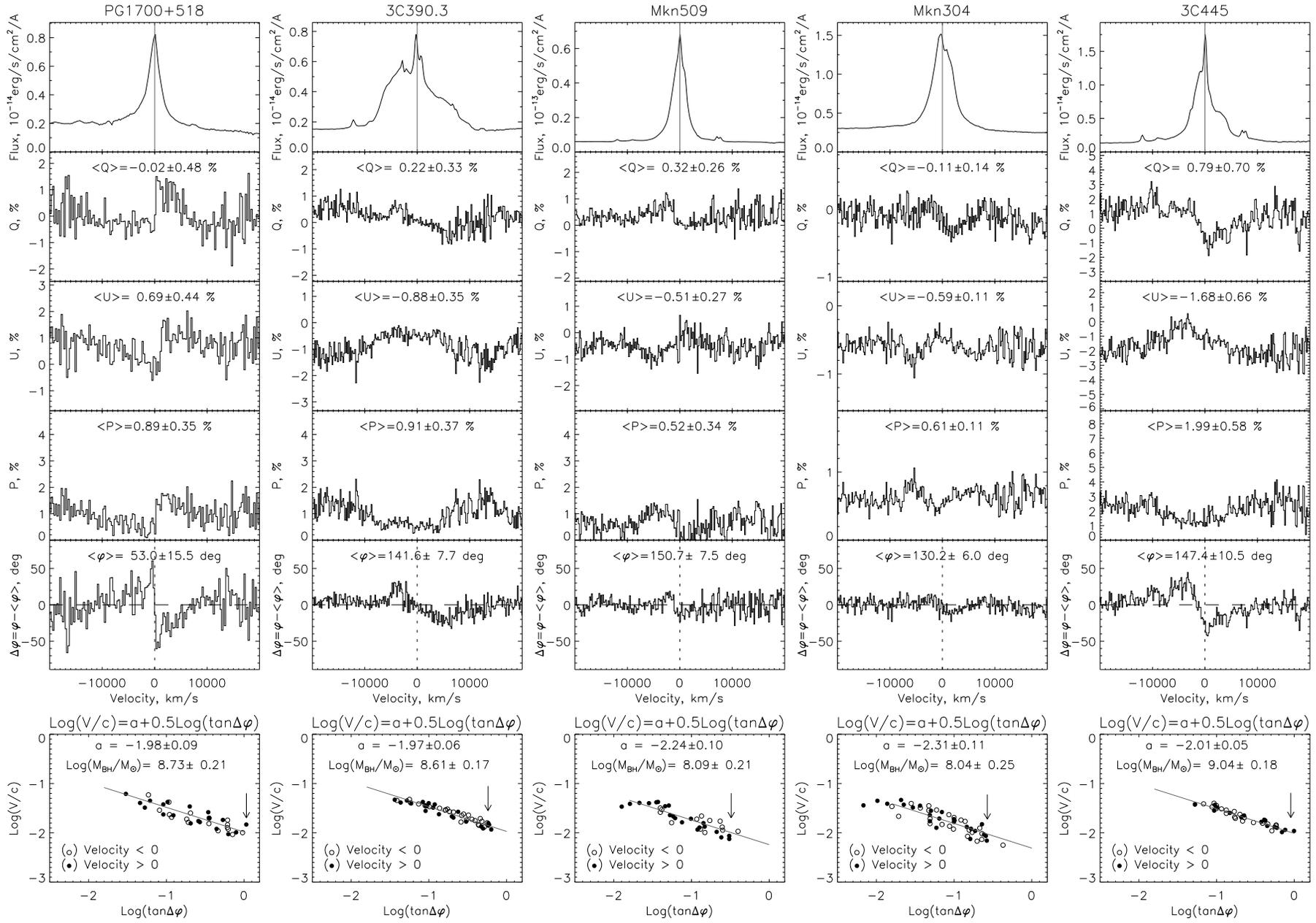}
\caption{The same as in Fig. \ref{res1}, but for PG 1700+518, 3C 390.3, Mkn 509, Mkn 304 and 3C 445.}
 \label{res6}
\end{figure}
\end{landscape}

\begin{figure}
\centering
\includegraphics{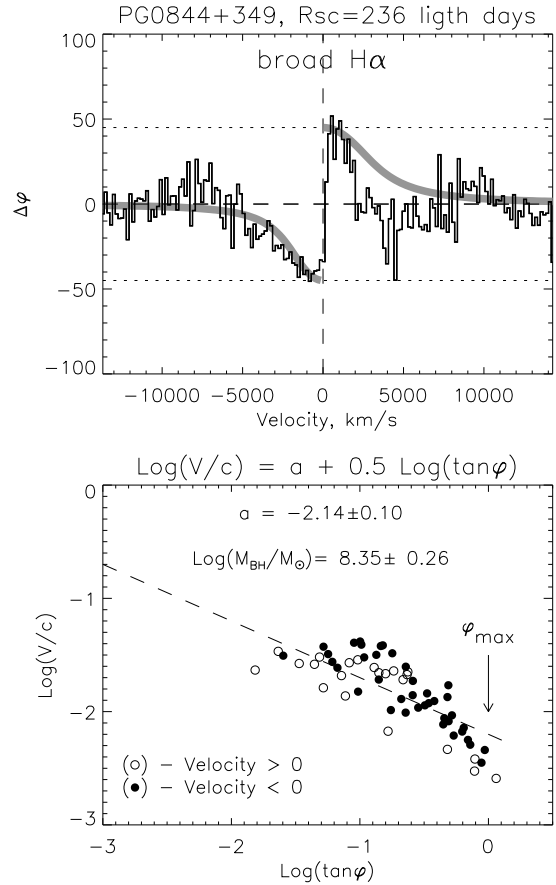}
\caption{The measured value of polarization across the line profile.
An example for PG0844+349, the panel up presents the horizontal S-shaped polarization
angle across the H$\alpha$ line profile, the measured (estimated) maximum P.A. is
shown on the plot. The panel bottom shows the  velocity vs. $\tan(\varphi)$
relation across line (in logarithmic scale).
} \label{angle-V}
\end{figure}

For each AGN from the sample in the corresponding bottom panel in Figs. \ref{res1}-\ref{res6}
 we give a trend between  $V/c$ and $\Delta\varphi$ (in the logarithmic scale) which indicates that in all AGNs from a  sample the equatorial scattering occurs and that Keplerian-like emission gas can be observed. The arrow on this  panels indicate the maximum  angle of  the polarization plane rotation across the H$\alpha$ line profile. The estimated masses are given in the bottom panel and also in Table \ref{tab3}.

As it can be seen in  bottom  panels Figs. \ref{res1}-\ref{res6}, relative change
the polarization angle  $\Delta\varphi$, mostly follow expected shape in the case of  Keplerian motion in the BLR. However, in some of the AGNs from the sample  $\Delta\varphi$  as a function of velocity has no clear  S-shape
 (see e.g.  of NGC4593, Mrk 231, Mrk 509. etc.), then we estimate the shape, as it is  shown in Fig. \ref{angle-V}, reproducing the shape assuming the Keplerian motion  \citep[assuming $b=0.5$ in Eq. (5), see][]{ap15}.
   Fig. \ref{angle-V} demonstrates an example of the analysis of
the observed change in the  polarization plane angle  in the
broad H$\alpha$ line for  PG0844+349.
The observed deviations from the expected dependence $\log(v/c) ${\it vs} $\log(\tan\varphi)$
are probably related to the presence of a non-circular motion in the BLR disc, as it was  found in the
galaxy Mkn6 \citep{af14}.

 Additionally, we carefully consider some objects where the changes across the H$\alpha$ broad  line are not prominent due to other influences,  e.g. in the case of 3C390.3 where  a strong depolarization was found  \citep[see][compare Fig. 1 from the paper with panel in  Fig. \ref{res6}]{af15}.

 In bottom panels of Figs. \ref{res1}-\ref{res6} the plots of Eq. (5) with
coefficient $a$ and $\log(M/M_\odot)$ (written on
the plots) are given. The full circles are taken from the red and
open ones from the blue broad H$\alpha$ side.

\subsubsection{The inner scattering radius}

The inner dusty torus radius ($R_{sc}$) can be estimated by the reverberation, i.e.
by finding the time lag between optical and infrared emission of an AGN
\citep[see e.g.][]{Kos14}.  Comparing the measurements with the luminosity,
it was found  that $R_{sc}\propto L^{-0.5}$ and  that the $R_{sc}$
seems to be  about twice as large as
the corresponding BLR radius \citep[][]{kis11}.

The relationship between $R_{sc}$ and luminosity  is expected for a case
where the dust temperature and the inner radius of the dusty torus are determined by
radiation equilibrium and sublimation of dust, respectively.
This dependence has been  confirmed by a
quantitatively estimated inner radius of the dust torus by taking into account
the wavelength dependent efficiency of dust grain absorption \citep{bar87}, where luminosity
at $\lambda=1216$\AA\ was used.

\begin{figure}
\centering
\includegraphics[width=8cm]{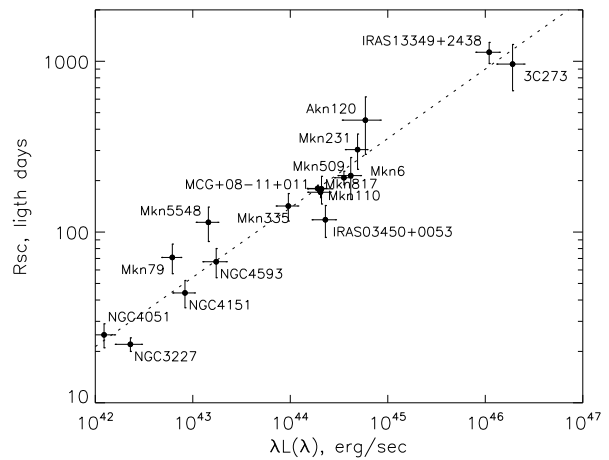}
\caption{The empirical dependence of the estimation of the inner
radius $R_{sc}$  of the torus on the UV-luminosity $\lambda L(\lambda)$ ~for  $\lambda=1516\AA$ .
The gray line shows the power  dependence with the power index of 0.421$\pm$0.026.
} \label{Rsc-UV}
\end{figure}

In Fig. \ref{Rsc-UV}
we show this relationship, which has been calculated
using the  empirical dependence of the inner radius of the torus on UV
luminosity. The correlation of the inner radius of the R$_{sc}$ torus with UV luminosity is physically justified and confirmed by measurements in the near IR emission. However, it should  be noted that the dependence of R$_{sc}$ -- UV flux in  literature
is confirmed by the extrapolation of the Spectral Energy Distribution (SED) at 1216 \AA\ and also  indirectly calculated from  the V-color fluxes.
To find the  {luminosity-radius} relationship, we used homogeneous GALEX data (F$_{UV}$ fluxes) measured at $\lambda$ 1516\AA.

%

The AB-magnitudes ($m_{AB}$) for AGNs from our sample   were taken from NASA/IPAC Extragalactic Database (NED)\footnote{https://ned.ipac.caltech.edu/}, and then have been converted to the flux using the well-known relationship
$\log F_{UV}=-0.4\cdot m_{AB}+7.33$.
The  absorption is taken into account as $A(F_{UV})=7.9\cdot E(B-V)$
\citep[see][]{gi07}, where $E(B-V)$ is the reddening.
For 2/3 of the objects from our sample  we could find in literature the  UV-fluxes  observed with   Hubble Space Telescope in  $\lambda$1450 \AA\ and $\lambda$1350\AA.
For a typical energy distribution of Sy1, the systematic difference between
the fluxes at $\lambda$1350\AA\ and $\lambda$1516 \AA\ is about 0.06 (in the
logarithmic scale). The standard deviation between our data measured at
$\lambda$1516\AA\ and  taken from the literature at $\lambda$1350 \AA\  and
$\lambda$1450 \AA\  \citep[][]{ve02,ka05} is in the
interval of 0.2-0.25.

As it can be seen in Fig. \ref{Rsc-UV} there is a good correlation between the $R_{sc}$ and luminosity at $\lambda=1516$\AA\  ($R_{sc}\propto L^{-m}$), where   the  power index $m$  is 0.421$\pm$0.026.  Using this calibration relation, we determined the  $R_{sc}$  for those objects for which we could  not find the $R_{sc}$ in literature and give these values for all objects in Table \ref{tab3}.

\section{ Analysis and discussion of the results}

\subsection{Polarization in the continuum}

Firstly we should note here that there is a wavelength dependent polarization in the continuum.
This cannot be due to the Thompson scattering in the disc, since it is  not wavelength dependent.
The change of the polarization as a function of wavelengths can be caused by three effects:
(i) The first mechanism is depolarization by Faraday rotation of the plane of polarization on the length of the
 electron free path in the magnetic accretion disc \citep[Faraday rotation depolarization, see e.g.][]{al96,gne97,al98}.
This mechanism gives polarization which depends on  wavelengths, the degree of polarization in this case
 is less than predicted by the radiation transfer mechanism in the accretion disk \citep [] [] {cha50};
(ii) The second mechanism is the scattering on the torus dust, where, as
it is well known, the degree of polarization is maximum at $\lambda_{max}=0.6$ mkm
\citep[see][]{se75}.
For $\lambda<\lambda_{max}$, the polarization is increasing with the wavelength, and for $\lambda>\lambda_{max}$
it is decreasing. The theoretical models show that the continuum polarization due to torus dust scattering is
dependent from the wavelengths \citep[][]{mar12}, and that for $\lambda<0.8$ mkm it is increasing;
 (iii) The third effect
is described in \cite{wo99}, where the observed wavelength dependence of the
 polarization in some AGNs can be explained by dust scattering in the optically thin cone. It is found (from modelling)
 that the multiple scattering in the cones is important for the optical depth larger than $\sim 0.1$.

It is difficult to separate effects giving the dependence of polarization  on the wavelength in continuum, but if we assume that magnetic field in the accretion disc is connected with the mass of central black hole \citep[magnetic coupling model, see e.g.][]{ma07}, then we can expect that the index of the  continuum slope is a function of the central BH mass, as it has been shown in \cite{af11}. We give  an empirical relation  between the index $n$ and $\log M_{BH}$.

\begin{figure}
\centering
\includegraphics[width=9cm]{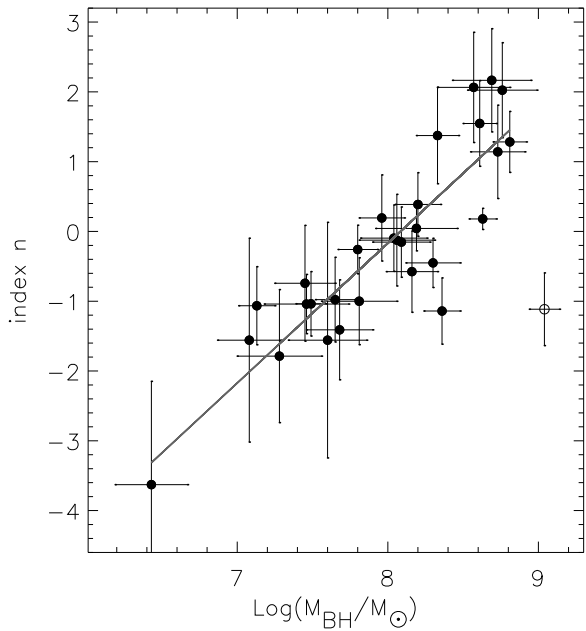}
\caption{Dependence of index $n$ onthe BH mass.
} \label{n-MBH}
\end{figure}

Using  the data given in  Tables \ref{cont-data} (for measured $n$)
and \ref{tab3} (for measured BH masses) we explored this relation
(see Fig. \ref{n-MBH}). As it can be seen in Fig. \ref{n-MBH},
only one object (3C445)   shows a strong deviation from the expected linear best fit (solid line).
The correlation coefficient for all points is 0.79. If we remove the outlier  (point representing
3C445) the correlation coefficient increases up to 0.88 for the BH mass range from $10^7 M\odot$ to $10^9M\odot$. The  statistical significance of this correlation is $< 10^{-3}$.
The slope of the obtained dependence is $1.6\pm0.3$, which is, within the error-bars, in an agreement  with the result obtained earlier by \cite{af11}. Taking that the Faraday rotation depolarization can give $n\sim \log M_{BH}$ relationship, one can conclude that it has an important
role in the continuum depolarization in Type 1 AGNs.

The obtained results show that probably the polarization in the continuum is mainly produced by  the accretion disc while the
 $n$ {\it vs}  $\log M$ relation points to    the magnetic field in accretion disc those dependent on the black hole mass \citep[][]{al96,al98}, that is expected
in the so called magnetic coupling model \citep[][]{ma07}. However, one can suppose that a small amount of the
polarized light in the continuum is caused by the scattering in the torus or cones, but it is hard to separate these
two mechanisms without detail modeling of polarization in the continuum.
The polarization of the most of Type 1 AGNs from our sample is between 0.5\% -- 1\% (see Table \ref{cont-data})
that is expected for broad line AGNs, however  there are four objects showing polarization higher than 2\% - IRAS 13349, Mkn 231, 3C 445 and Mkn 704.


\begin{table*}

\begin{center}
\small
\caption[]{The polarization parameters for H$\alpha$. From left to the right:
name of object, logarithm of the UV-luminosity at $\lambda=$1516\AA, corresponding
reference, $R_{sc}$ - the inner radius of torus, corresponding reference and logarithm of the BH masses (measured by polarization method).
}\label{tab3}

\begin{tabular}{lccccc}

\hline \hline
\\
Object&$Log({\lambda}L_{\lambda})$&Ref.&$R_{sc}$&Ref.& $Log\bigl({M_{BH}\over{M_{\odot}}}\bigr)$ \\
      &$\lambda=1516\AA$ &         & ligth day   &    &                                           \\

\hline

Mkn335		    &43.98$\pm$0.25& 4,5	&142$\pm$26&2&7.49$\pm$0.25\\
Mkn1501		    &44.06$\pm$0.21& 4,5	&398$\pm$46&1&8.57$\pm$0.26\\
Mkn1148		    &44.54$\pm$0.20& 4	    &232$\pm$28&1&8.69$\pm$0.18\\
1Zw1		    &44.21$\pm$0.20& 4,5	& 93$\pm$9 &2&7.46$\pm$0.30\\
IRAS03450+0055  &44.36$\pm$0.27& 4	    &118$\pm$25 &1&8.06$\pm$0.31\\
3C120		    &44.57$\pm$0.29& 4,5,6 	&223$\pm$37&1&7.68$\pm$0.26\\
Akn120		    &44.77$\pm$0.42& 4,5,6 	&~452$\pm$167&3&8.33$\pm$0.16\\
MCG+08-11+011   &44.32$\pm$0.28& 4      &179$\pm$33&2&8.19$\pm$0.32\\
Mkn6		    &44.62$\pm$0.26& 4	    &214$\pm$59&3&8.20$\pm$0.29\\
Mkn79		    &42.79$\pm$0.22& 4,5,6	&~71$\pm$14	&3&7.45$\pm$0.27\\
PG0844+349      &44.56$\pm$0.25& 4,5    &236$\pm$25	&1&8.35$\pm$0.26\\
Mkn704		    &43.62$\pm$0.28& 4 	    &~~98$\pm$13  &1&7.80$\pm$0.17\\
Mkn110		    &44.31$\pm$0.26& 4,5,6	&171$\pm$12	&3&8.32$\pm$0.21\\
NGC3227		    &42.36$\pm$0.30& 4,6 	&22$\pm$2 	    &2&7.31$\pm$0.22\\
NGC4151		    &42.92$\pm$0.25& 4,5,6 	&44$\pm$8 	    &3&7.08$\pm$0.27\\
NGC4051		    &42.09$\pm$0.28& 4,6	&25$\pm$4	    &2&6.43$\pm$0.29\\
3C273		    &46.28$\pm$0.31& 4,5	&~963$\pm$291	&3&8.81$\pm$0.24\\
NGC4593		    &43.24$\pm$0.27& 4,6	&~67$\pm$13	    &2&7.28$\pm$0.19\\
Mkn231		    &44.69$\pm$0.25& 4	    &304$\pm$71	&2&8.30$\pm$0.27\\
IRAS13349+2438  &46.04$\pm$0.27& 4      &1130$\pm$160	&3&8.63$\pm$0.15\\
Mkn668		    &42.95$\pm$0.23& 4	    &53$\pm$5	    &1&8.16$\pm$0.19\\
NGC5548		    &43.16$\pm$0.25& 4,5,6 	&114$\pm$26	&2&7.81$\pm$0.32\\
Mkn817		    &44.28$\pm$0.22& 4	    &180$\pm$18	&2& 7.65$\pm$0.16\\
Mkn841		    &44.29$\pm$0.29& 4,5	&182$\pm$28	&1&8.76$\pm$0.27\\
Mkn876		    &44.65$\pm$0.27& 4,5,6 	&254$\pm$39	&1&8.36$\pm$0.18\\
PG1700+518	    &44.73$\pm$0.20& 4,6 	&277$\pm$34	&1&8.73$\pm$0.21\\
3C390.3		    &44.38$\pm$0.27& 4,5,6	&200$\pm$29	&1&8.61$\pm$0.17\\
Mkn509		    &44.55$\pm$0.20& 4,5,6	&208$\pm$19	&3&8.09$\pm$0.21\\
Mkn304		    &44.64$\pm$0.24& 4	    &255$\pm$35	&1&8.04$\pm$0.25\\
3C445		    &45.64$\pm$0.28& 4	    &~650$\pm$115	&1&9.04$\pm$0.18\\

\hline

\end{tabular}
\\

{References:  1 - this work, 2 - Koshida at al. (2014),
3 - Kishimoto et al. (2011), 4 - GALEX , 5 - Vestergaard (2002), 6 - Kaspi t al. (2005) \\}
\end{center}
\end{table*}

\subsection{Measurements of BH mass from H$\alpha$ polarization}

We used the method given in \cite{ap15} to measure  BH masses. { The method does not depend of the inclination of the BLR in the case equatorial scattering, and detailed discussion} of the method is given in \cite{sav17}, where the equatorial polarization in the broad lines has been modelled with STOKES code. It seems that the method gives reasonable accurate estimates of BH masses, even if in the BLR  some kind of outflows and inflows are present in addition to Keplerian motion, \citep[see in more details in][]{sav17}. Also, it was shown in \cite{ap15} that measured BH masses from polarization are in a good agreement with ones obtained by reverberation.

Our measurements of  BH masses are presented in Table \ref{tab3}. As it was shown in  \cite{tre02} there is a relation
between the BH masses and host galaxy bulge  stellar velocity dispersion ($\sigma_*$) that follows  $M_{BH}\sim \sigma_*^4$. This  is caused by well known connection between the brightness of the bulge and   stellar velocity dispersion \citep[][]{fab76}.
We explore this relationship using our BH mass measurements and
stellar velocity dispersion given in  \cite{on04}.

\begin{figure}
\centering
\includegraphics[width=8cm]{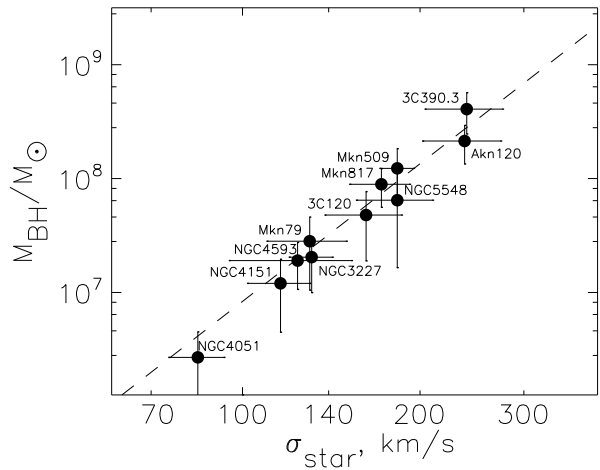}
\caption{Our measurements of  BH masses as a function of host galaxy bulge
stellar velocity dispersion $\sigma_*$ \citep[taken from][]{on04}. 
The dashed line shows the dependence of
 $M_{BH}\sim\sigma_*^4$ taken from \citet{tre02}.} 
 \label{fig5}
\end{figure}

As it can be seen in Fig. \ref{fig5}, the relation between our measurements of BH masses from
polarization
and   stellar velocity dispersions follow expected one.
The correlation between BH mass and $\sigma$ is very  high (r=0.98).

Our estimates of the BH masses and the  relation between $M_{BH}$ and $\sigma$  are in a  good agreement with those given by \cite{tre02} and in the future the BH mass measured from polarization in the broad lines can be used for calibration purposes of other methods for BH mass estimation.

We compared the measurements of BH mass given in Table 3 with our previous results by \cite{ap15}.  It turned out that the latest data have smaller errors than the previous ones,  due to our new of the method of simultaneous measurement of Stokes parameters.

\subsection{The BLR inclinations}

In the case of the reverberation BH mass measurements (also using the relations for single-epoch mass measurements) it is very
important to have information about the disc inclination \citep[see][]{col06}. In the polarization method, the  BH mass estimates
do not depend on the BLR inclination. Consequently, comparing masses obtained by reverberation and those obtained by polarization method,
one can measure the BLR inclinations.

Using the reverberation method, BH masses can be obtained from following relation
\citep[see][]{pet14}:

$$M_{BH}=f{R_{BLR} \sigma_V^2\over G}=f\cdot VP,\eqno(6)$$
where $R_{BLR}$ is the photometric size of the BLR obtained
by the reverberation, and $\sigma_V$ is the corresponding  orbital velocity which can
be estimated from the broad lines.
The virial factor $f$ depends on the inclination
and geometry of the BLR, and for Keplerian motion $f=1$. However, from the comparison of
reverberation and  stellar velocity dispersion BH
mass estimates it is obtained that $f>1$. As we noted above, the role of additional gas
motion to the Keplerian one can affect the
coefficient $f$. Since we measure the BH mass from polarization, and the Keplerian motion is dominant (can be explored by a relationship between velocities and $\tan\varphi$), in our sample the $f$ the most affected by BLR  inclination. The measured
velocities (from the width of lines) depend on inclination
as ~ $\sigma_V=\sigma_{V_{obs}}/\sin i$,
therefore we have that $f=1/\sin^2i$.

\begin{table*}
\begin{center}
\caption[]{The measured BLR inclinations. From the left to the right are given: object name, luminosity, references for luminosity,
estimated radius of the BLR, references for $R_{BLR}$ estimates, dispersion obtained from FWHM, virial product divided with
 by the BH mass, BH mass, inclination of the BLR,
and maximal size of the BLR.
}\label{tab4}

\begin{tabular}{lcccccccccc}

\hline \hline
\\
Object&$Log({\lambda}L_{\lambda})$ & Ref.& $R_{BLR}(H\beta)$ & Ref. & $\sigma_{V}(H\beta) $  & $Log\bigl({VP\over{M_{\odot}}}\bigr)$ & $Log\bigl({M_{BH}\over{M_{\odot}}}\bigr)$ & $i_{BLR}$ & $R_{max}$\\

 &  $\lambda=5100\AA$ &  &ligth day&  & km/s &  &  & deg.& ligth days  \\
\hline
Mkn335          &43.71$\pm$0.06&  2& 15.7$\pm$ 3.4&  2&2253$\pm$ 85&7.19$\pm$0.13&7.49$\pm$0.25&44.5$\pm$ 9.0& 119$\pm$ 17\\
Mkn1501         &44.41$\pm$0.07&  1& 72.3$\pm$ 5.4&  1&2385$\pm$ 83&7.90$\pm$0.06&8.57$\pm$0.26&28.8$\pm$ 6.3& 239$\pm$ 18\\
Mkn1148         &44.00$\pm$0.01&  1& 34.3$\pm$ 0.1&  1&2006$\pm$150&7.43$\pm$0.07&8.69$\pm$0.18&13.7$\pm$ 1.4& 156$\pm$ 11\\
1Zw1            &43.49$\pm$0.06&  1& 18.7$\pm$ 2.5&  1&2128$\pm$119&7.22$\pm$0.11&7.46$\pm$0.30&45.4$\pm$11.2&  19$\pm$  5\\
IRAS03450+0055  &43.77$\pm$0.05&  1& 27.3$\pm$ 2.7&  1&1766$\pm$136&7.22$\pm$0.11&8.06$\pm$0.31&23.3$\pm$ 5.5& 102$\pm$  7\\
3C120           &43.87$\pm$0.05&  2& 25.6$\pm$ 2.9&  2&2371$\pm$ 87&7.45$\pm$0.08&7.68$\pm$0.26&50.7$\pm$12.4& 207$\pm$ 14\\
Akn120          &43.78$\pm$0.07&  2& 39.7$\pm$ 3.0&  2&2543$\pm$ 33&7.70$\pm$0.04&8.33$\pm$0.16&29.3$\pm$ 3.9& 220$\pm$ 24\\
MCG+08-11-011   &43.59$\pm$0.08&  4& 15.0$\pm$ 0.3&  4&2186$\pm$ 44&7.14$\pm$0.03&8.19$\pm$0.32&18.1$\pm$ 4.0&  79$\pm$ 12\\
Mrk6            &43.66$\pm$0.05&  1& 20.6$\pm$ 2.0&  6&3351$\pm$113&7.65$\pm$0.07&8.20$\pm$0.29&32.9$\pm$ 5.0& 118$\pm$ 15\\
Mrk79           &43.61$\pm$0.04&  2& 15.2$\pm$ 4.0&  2&2456$\pm$ 83&7.25$\pm$0.15&7.45$\pm$0.27&49.1$\pm$ 9.8&  37$\pm$  8\\
PG0844+349      &44.24$\pm$0.04&  2& 32.3$\pm$13.0&  2&2737$\pm$106&7.67$\pm$0.23&8.35$\pm$0.26&29.6$\pm$ 8.8& 212$\pm$ 14\\
Mkn704          &43.39$\pm$0.02&  1& 16.5$\pm$ 2.0&  1&2742$\pm$ 45&7.38$\pm$0.07&7.80$\pm$0.17&39.0$\pm$ 5.5&  29$\pm$  3\\
Mkn110          &43.60$\pm$0.04&  2& 25.5$\pm$ 5.0&  2&1953$\pm$107&7.28$\pm$0.14&8.32$\pm$0.21&18.0$\pm$ 3.1&  27$\pm$  7\\
NGC3227         &42.24$\pm$0.11&  2&  7.8$\pm$ 4.0&  2&1881$\pm$ 96&6.73$\pm$0.31&7.31$\pm$0.22&31.4$\pm$10.3&  11$\pm$  6\\
NGC4151         &42.09$\pm$0.22&  2&  6.6$\pm$ 0.1&  2&2289$\pm$ 92&6.83$\pm$0.04&7.08$\pm$0.27&53.4$\pm$14.8&  24$\pm$  7\\
NGC4051         &41.96$\pm$0.20&  7&  2.5$\pm$ 1.0&  8&1146$\pm$ 88&5.81$\pm$0.27&6.43$\pm$0.29&32.1$\pm$10.4&  14$\pm$  8\\
3C273           &45.90$\pm$0.02&  7&306.8$\pm$90.9&  9&1777$\pm$150&8.28$\pm$0.22&8.81$\pm$0.24&34.3$\pm$ 7.8& 837$\pm$ 67\\
NGC4593         &42.87$\pm$0.18&  7&  4.5$\pm$ 0.6&  5&2146$\pm$ 59&6.61$\pm$0.09&7.28$\pm$0.19&27.7$\pm$ 2.7&  23$\pm$  6\\
Mkn231          &44.35$\pm$0.05&  1& 51.2$\pm$ 5.9&  1&2777$\pm$ 93&7.89$\pm$0.08&8.30$\pm$0.27&39.8$\pm$ 7.7& 110$\pm$ 16\\
IRAS13349+2438  &44.72$\pm$0.06&  1& 81.2$\pm$11.2&  1&2971$\pm$ 93&8.15$\pm$0.09&8.63$\pm$0.15&35.3$\pm$ 3.9& 138$\pm$ 41\\
Mkn668          &43.33$\pm$0.09&  1& 15.4$\pm$ 6.6&  1&3357$\pm$ 99&7.53$\pm$0.23&8.16$\pm$0.19&30.6$\pm$ 7.6&  47$\pm$  3\\
NGC5548         &43.23$\pm$0.10&  2&  8.7$\pm$ 0.5&  9&3646$\pm$105&7.35$\pm$0.05&7.81$\pm$0.32&38.3$\pm$ 9.4&  43$\pm$  8\\
Mkn817          &43.68$\pm$0.05&  2& 21.2$\pm$ 4.7&  2&2537$\pm$ 61&7.43$\pm$0.12&7.65$\pm$0.16&49.6$\pm$ 7.4&  202$\pm$ 27\\
Mkn841          &44.29$\pm$0.11&  1& 48.6$\pm$12.3&  1&2629$\pm$151&7.82$\pm$0.17&8.76$\pm$0.27&20.6$\pm$ 4.8& 105$\pm$  8\\
Mkn876          &44.71$\pm$0.03&  2& 35.0$\pm$15.1& 10&2944$\pm$ 44&7.77$\pm$0.22&8.36$\pm$0.18&31.9$\pm$ 7.1& 184$\pm$ 13\\
PG1700+518      &45.53$\pm$0.02&  2&251.8$\pm$42.4& 10&3405$\pm$136&8.76$\pm$0.11&8.73$\pm$0.21&60.3$\pm$ 8.1& 297$\pm$ 20\\
3C390.3         &43.82$\pm$0.02&  2& 23.6$\pm$ 5.4&  2&5424$\pm$185&8.13$\pm$0.13&8.61$\pm$0.17&36.0$\pm$ 5.5& 115$\pm$  9\\
Mkn509          &44.16$\pm$0.10&  2& 79.6$\pm$ 4.0&  2&1843$\pm$ 43&7.72$\pm$0.04&8.09$\pm$0.21&42.5$\pm$ 8.4&  67$\pm$ 11\\
Mkn304          &44.38$\pm$0.13&  1& 54.1$\pm$16.2&  1&2307$\pm$ 37&7.75$\pm$0.15&8.04$\pm$0.25&45.2$\pm$10.3&  68$\pm$  9\\
3C445           &44.92$\pm$0.05&  1&103.2$\pm$11.9&  1&2513$\pm$163&8.10$\pm$0.11&9.04$\pm$0.18&20.1$\pm$ 2.3& 585$\pm$ 41\\

\hline
\end{tabular}
\\
{{\bf References}:  1 - this work, 2 - Bentz et al. (2009), 3 - Fausnaugh (2017), 4 - Zu et al. (2011),
5 - Doroshenko  et al. (2012), 5 -  Bentz et al. (2013), 7 - Denney et al. (2009),
8 - Peterson et al. (2004), 9 - Kaspi et al. (2000)\\}
\end{center}
\end{table*}

$VP$ in Eq. (6) is the so-called virial product
$$VP={R_{BLR}\sigma_V^2\over G}$$
 and can be used for the BLR inclination
measurements. If we estimate the $R_{BLR}$ from reverberation, and we measure the
dispersion velocity assuming the Gaussian profiles, we are able to obtain the BLR inclination as
$$\sin^2i={VP\over M_{BH}},\eqno(7)$$
where $M_{BH}$ is the mass measured by polarization from the
H$\alpha$ broad line (taken from Table \ref{tab3}).

In literature we can find the relationships between the $R_{BLR}$ and luminosity in the continuum
\citep[see e.g.][etc]{ka05,vp06,on08,wa09,tn12,ti13,me16,co17}.

Evaluation of the width of the broad lines cited in the literature can vary by a factor of 2-2.5 times and depends on the applied methodology \footnote{http://www.astro.gsu.edu/AGNmass/}. To obtain homogeneous estimates $\sigma_V$ in Table \ref{tab4}, we measured FWHM of a broad  $H\beta$ line on our spectra, which in the case of the Gauss profile is $2\sqrt{2\ln2}\sigma_V\approx 2.355 \sigma_V$.

\begin{figure}
\centering
\includegraphics[width=8cm]{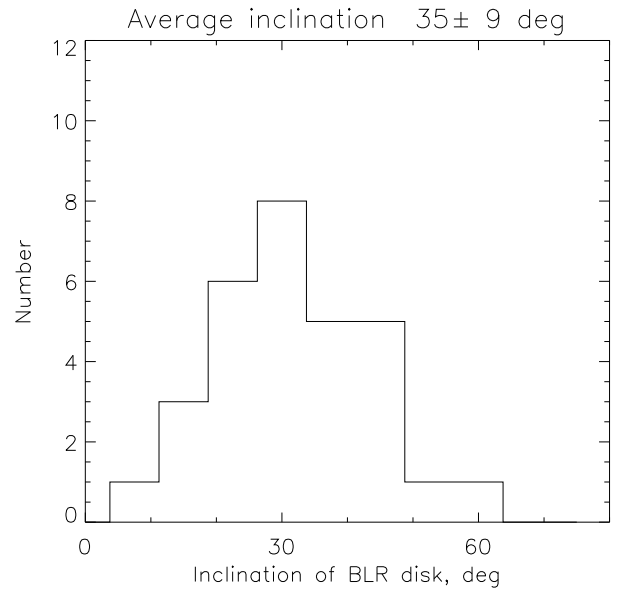}
\caption{Distribution of inclination BLR disc in our sample.}
 \label{fig6}
\end{figure}

 In Table \ref{tab4} we give the parameters
from the literature that was used to determine the
BLR size and also the estimated BLR inclination  values using Eq. (7).
In Fig. \ref{fig6} we show the histogram of the inclination, where can it be seen
that the BLR inclination is mostly  in the range between $\sim 20^\circ$ and
$\sim 40^\circ$. The  average BLR inclination of the sample is $35\pm9$ degrees.
 This is in a good agreement with  numerical simulation provided by  \cite{sav17}, where
the BLR inclinations are between $25^\circ$ and $45^\circ$.

  It should be mentioned  that
\cite{kol03} first compared the BH masses obtained by reverberation method with ones
that have been estimated using the gravitational redshift method to find the inclination
of $\sim19^\circ$   for the Mrk 110 BLR.  Also there are several estimations
of the BLR inclinations obtained by fitting of the broad double peaked lines with an accretion disc model, where
inclinations seem to be between $\sim 20^\circ$ and $\sim 40^\circ$  \citep[see e.g.][etc.]{er94,er96,pop04,bon09,sb17}, that
 fits the results we obtained from our method very well.

\subsection{Maximal dimensions of the disc-like BLR}

Taking into account that the main polarization mechanism in broad lines is the equatorial scattering, the maximal
radius of the disc-like BLR ($R_{max}$) is connected with the inner part of the
scattering region as \citep[see Fig. 1 in][]{ap15}
$$R_{max}=R_{sc}\tan(\varphi_{max}),$$
where $\varphi_{max}$ is the maximum angle of the polarization line across a broad line.

 Our estimates of  $R_{max}$ are given in the last column of Table \ref{tab4},
and the $\varphi_{max}$ is marked  with  arrows   for each object in Figs.
\ref{res1}-\ref{res6}. Using data for $R_{sc}$ and  $R_{BLR}$ given in Tables
\ref{tab3}  and \ref{tab4}, we obtained that the averaged ratio $R_{sc}/R_{BLR}\sim 1.72\pm0.48$
that is  in a good agreement with  numerical model given in \cite{sav17}, where this
ratio is considered to be in a  range of 1.5-2.5.

\begin{figure}
\centering
\includegraphics[width=8cm]{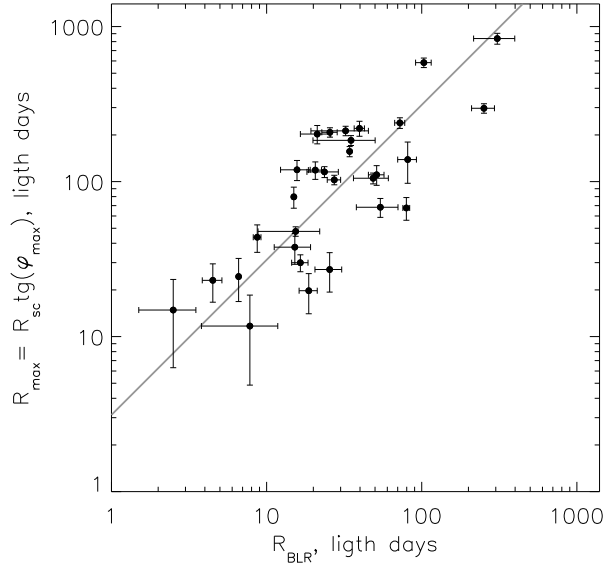}
\caption{The  maximal ($R_{max}$) {\it vs} photometric BLR ($R_{BLR}$) radius.  
The solid line represents the best fit.}
\label{fig7}
\end{figure}

One can assume that $R_{max}\sim R_{BLR}$,  i.e. that $R_{max}$ coincides with the
photometric BLR radius. In this case  there should be a good correlation between the sizes of the BLR given in Table \ref{tab4} and
the $R_{max}$  calculated by the equation above. In Fig. \ref{fig7} we show the relation between these
two radii, being in is a very good correlation (r=0.84),
the photometric BLR size depends on the  maximum  BLR radius as
$$R_{BLR}=(0.31\pm 0.17)R_{max}.$$

To explain this relation, let us consider the connection between photometric
size of the BLR and its real size taking   photoionization as the  mechanism for the broad line emission.
Admitting  that the reverberation gives the photometic size of the BLR, one can write:

$$R_{BLR}=c\tau\approx{\int^{R_{max}}_{R_{min}}{I(r)r}dr\over {\int^{R_{max}}_{R_{min}}{I(r)}dr}},$$
where $R_{min}$ and $R_{max}$ are the inner and outer radii of the BLR, respectively. $\tau$ is the time lag obtained from
reverberation and $c$ is the light speed, and  $I(r)$ is the disc emissivity. If we accept that the emissivity of
the disc is $I(r)\sim r^\alpha$ for $\alpha\neq -1$,  assuming that $R_{max}>>R_{min}$  then we have the relation
between the photometric and maximum BLR radius:

$$R_{BLR}\approx {1+\alpha\over{2+\alpha}}R_{max},$$
that in the case of the flat-brightness disc ($\alpha=0$) gives $R_{BLR}=0.5R_{max}$, but if we take more realible values
$\alpha=-3/4$ \citep[][]{ss73} the connection is $R_{BLR}=0.2 R_{max}$. We obtain the relation that is  between
these two values, and corresponds to
$\alpha\approx -0.57$, that is flatter than one expected for the  classical disc emission
coefficient $\alpha=-0.75$.

\section{Conclusions}

Here we  presented results of the spectropolarimetric observations of 30 Type 1 AGNs. The measured polarization properties
have been used for exploring polarization mechanisms in the continuum and determination of the BH masses and the parameters of the  BLR
(inclination and emissivity) using polarization in the broad H$\alpha$ line.

According to  of our analysis of polarization observations of the sample of 30 broad line AGNs,
we can outline following conclusions:

(i) The continuum polarization in the sample of 30 Type 1 AGNs is on the level of 1\%, that is expected for this
class of AGNs \citep[see e.g.][]{be90}.
The continuum polarization for  all AGNs from the sample is wavelength dependent, indicating
some effect of depolarization by the magnetic field or other effects. Additionally,
the degree of polarization is smaller than one expected due to Thompson scattering
 in an accretion disc. Also the index of the polarized continuum slope is in a relatively good correlation with the
black hole masses. This indicates that probably Faraday rotation has a significant role in the depolarization of the continuum light
emitted from a magnetized disc.

(ii) The polarization in the H$\alpha$ broad line shows that the
equatorial scattering is a dominant mechanism of polarization in the broad lines in
Type 1 AGNs. The polarization angle across the broad line profile indicates dominant Keplerian-like motion in the BLR. Using the method
given in \cite{ap15} we estimated the masses of central BHs, and found that they follow expected $M-\sigma$ relation. Also
we confirmed that BH masses obtained from  polarization can be used as calibration masses for other methods.

(iii) Using BH masses obtained by polarization method and corresponding 'virial products' we  estimated the  BLR inclinations in  our sample. We found that
the most of them have  the BLR inclination between 20 and 40 degrees with an
average BLR inclination { of  $\sim$ 35 degrees}.

(iv) The coefficient of  BLR emissivity seems to be smaller than one expected in the Shakura-Sunyaev accretion disc ($\alpha\sim-0.75$) and the BLR
tends to have a flatter emissivity ($\alpha\sim-0.57$)
than the accretion disc, this should be taken into account in modelling the AGN broad line profiles.

\section*{Acknowledgements}
The results of observations were obtained with the 6-m BTA telescope
of the Special Astrophysical Observatory of Academy of Sciences,
operating with the financial support of the Ministry of Education
and Science of Russian Federation (state contracts no.
16.552.11.7028, 16.518.11.7073).The authors also express appreciation
to the Large Telescope Program Committee of the RAS for
the possibility of implementing the program of Spectropolarimetric
observations at the BTA. This work was supported by the Russian
Foundation for Basic Research ({project N15-02-02101}) and
the Ministry of Education, Science and Technological Development (Republic of Serbia) through
the project Astrophysical Spectroscopy of Extragalactic Objects
(176001).  We are grateful  to M. Gabdeeev for his help in spectropolarometric observations and also
we would like to thank to  referee for very useful comments.

\end{document}